\documentclass[a4paper]{jpconf}
\usepackage{graphicx}
\usepackage[utf8]{inputenc}
\usepackage[left=2.55cm, right=2.55cm, top=2.55cm, bottom=2.55cm]{geometry}
\usepackage{amsmath,amssymb,amsbsy}
\usepackage{slashed}
\usepackage{xcolor}
\usepackage{graphicx}
\usepackage{url}
\usepackage{cancel}
\usepackage{cite}
\usepackage[colorlinks=true,allcolors=darkpurple,pdfborder={0 0 0},linktocpage=false]{hyperref}
\usepackage{tabularx,booktabs}
\usepackage{multicol}
\usepackage{units}
\usepackage{xspace}
\usepackage[labelfont=bf]{caption}
\usepackage[section]{placeins}
\usepackage{hyperref}

\definecolor{darkred}{rgb}{0.6,0,0}
\definecolor{darkpurple}{rgb}{0.5,0,0.5}
\def\vev#1{\left\langle #1\right\rangle}
\def\hc{\text{h.c.}}

\def\neta{n_\eta}
\def\nN{n_N}
\def\Tr{\text{Tr}}

\begin{document}
\title{A generalization of the Scotogenic model}

\author{Pablo Escribano}

\address{Instituto de F\'{\i}sica Corpuscular (CSIC-Universitat de Val\`{e}ncia), \\
C/ Catedr\'atico Jos\'e Beltr\'an 2, E-46980 Paterna (Valencia), Spain}

\ead{pablo.escribano@ific.uv.es}

\begin{abstract}
The Scotogenic model is a radiative neutrino mass model able to induce Majorana neutrino masses at the 1-loop level and simultaneously include a dark matter candidate. In this work, we generalize the original Scotogenic model to arbitrary numbers of generations of the Scotogenic states. After that, we present the light neutrino mass matrix, with some details of its derivation, and provide a useful approximate expression as well. Finally, we numerically solve the Renormalization Group Equations to explore the high-energy behavior of the model.~\protect\footnote{This work is based on a talk given at the TAUP 2021 Conference and a video of it can be found on \href{https://www.youtube.com/watch?v=OHq-6Nerk8g\&t=1s}{this link}.}
\end{abstract}

\section{Introduction}
\label{sec:intro}

The Scotogenic model \cite{Ma:2006km} is one of the most popular radiative neutrono mass models. With only the addition of three singlet fermions, one scalar doublet and a \textit{dark} $\mathbb{Z}_2$ symmetry under which the new particles are odd, it is able to account for neutrino masses at the 1-loop level and also to obtain a dark matter (DM) candidate. Although many variations and extensions of the original model have been presented since its appearance, our aim here is to introduce the \textit{general Scotogenic model}~\cite{Escribano:2020iqq}, which includes arbitrary numbers of generations of the Scotogenic states.

\section{The general Scotogenic model}
\label{sec:scot}

In our generalization of the Scotogenic model, we extend the SM by unspecified numbers, $\nN$ and $\neta$, of singlet fermions ($N$) and inert scalar doublets ($\eta$), respectively. Using the $(\nN,\neta)$ values, one can label particular cases of this particle spectrum. In addition, exactly as in the original Scotogenic model~\cite{Ma:2006km}, we enlarge the symmetry group of the SM with a \emph{dark} $\mathbb{Z}_2$ parity. Under this symmetry, all the new particles are odd, while the SM states are even.

For our discussion, the relevant Yukawa and bare mass terms are 
\begin{equation} \label{eq:yukawa}
\mathcal{L}_N \supset y_{n a \alpha} \, \overline{N}_n \, \eta_a \, \ell_L^\alpha + \frac{1}{2} \, M_{N_n} \, \overline{N^c}_n \, N_n + \hc \, ,
\end{equation}
where $n=1,\dots,\nN$, $a=1,\dots,\neta$ and $\alpha=1,2,3$ are
generation indices and $M_N$ has been chosen diagonal without lossing generality. Regarding the scalar potential, it can be written as
\begin{equation} 
  \begin{split}
    \mathcal{V} &=m_{H}^{2} H^{\dagger} H+\left(m_{\eta}^{2}\right)_{a b} \eta_{a}^{\dagger} \eta_{b}+\frac{1}{2} \, \lambda_{1}\left(H^{\dagger} H\right)^{2}+\frac{1}{2} \, \lambda_{2}^{a b c d}\left(\eta_{a}^{\dagger} \eta_{b}\right)\left(\eta_{c}^{\dagger} \eta_{d}\right) \\ 
    &+\lambda_{3}^{a b}\left(H^{\dagger} H\right)\left(\eta_{a}^{\dagger} \eta_{b}\right)+\lambda_{4}^{a b}\left(H^{\dagger} \eta_{a}\right)\left(\eta_{b}^{\dagger} H\right) +\frac{1}{2} \left[\lambda_{5}^{a b} \left(H^{\dagger} \eta_{a}\right)\left(H^{\dagger} \eta_{b}\right) + \, \hc \right] \, ,
  \end{split}
\label{eq:potential}
\end{equation}
where all the indices are $\eta$ generation indices. Again, $m_\eta^2$ is considered to be diagonal without loss of generality and $\lambda_{3,4}$ must be Hermitian. Finally, $\lambda_5$ must be symmetric and its presence will play a major role in the generation of neutrino masses.

Assuming that minimizing the scalar potential in Eq. \eqref{eq:potential} leads to this vacuum configuration: $\vev{H^0} = \frac{v}{\sqrt{2}}$ , $\vev{\eta_a^0} = 0$, with $a = 1, \dots, \neta$, only the neutral component of $H$ acquires a non-zero vacuum expectation value (VEV), breaking the electroweak symmetry. This way, the $\mathbb{Z}_2$ symmetry remains unbroken, ensuring the lightest $\mathbb{Z}_2$-charged particle stability.

We will assume in the following that the parameters in the scalar potential are all real. Therefore, CP will be conserved in the scalar sector. This way, the real and imaginary components of $\eta_a^0$,
\begin{equation}
	\eta_{a}^{0}=\frac{1}{\sqrt{2}} \, \left(\eta_{R_{a}} + i \, \eta_{I_{a}} \right) \, .
\end{equation}
do not mix. Note that $\eta_{A_a}$, where $A = R, I$, are gauge eigenstates and they are related to the mass eigenstates, $\hat{\eta}_{A_b}$, by $\eta_A = V_A \, \hat{\eta}_A$, where, since the scalar parameters are real, the $V_{A}$ matrices are orthogonal.

Now, since the analytical expressions for the scalar masses and the mixing matrices include non-trivial combinations of the parameters in the scalar potential, it proves convenient to work under the assumptions
\begin{equation} \label{eq:assumptions}
\lambda_{3,4}^{a a} \, \frac{v^2}{2} \ll \left(m_{\eta}^{2}\right)_{a a} \quad \text{and} \quad \lambda_{5}^{a b} \ll \lambda_{3,4}^{a b} \ll 1 \, ,
\end{equation}
%
which are technically natural~\cite{tHooft:1979rat}. Then, once the electroweak symmetry gets broken and using the previous assumptions, one can find simple expressions for the $m_{A_{a}}^{2}$ mass eigenvalues,
\begin{equation}
  m_{R_{a}}^{2} = \left(m_{\eta}^{2}\right)_{a a} + \left( \lambda_{3}^{a a} + \lambda_{4}^{a a} + \lambda_{5}^{a a} \right) \, \frac{v^2}{2} \quad , \quad m_{I_{a}}^{2} = \left(m_{\eta}^{2}\right)_{a a} + \left( \lambda_{3}^{a a} + \lambda_{4}^{a a} - \lambda_{5}^{a a} \right) \, \frac{v^2}{2} \, ,
\end{equation}
where we note that their difference vanishes in the limit $\lambda_5 \to 0$. Concerning the $V_A$ matrices, they can be written as a product of $n_\eta (n_\eta - 1)/2$ rotation matrices,
whose scalar mixing angles are given by
\begin{equation} \label{eq:angles}
\tan 2 \, \theta_A^{ab} = \frac{2 \, (\mathcal{M}_{A}^{2})_{a b}}{(\mathcal{M}_{A}^{2})_{b b} - (\mathcal{M}_{A}^{2})_{a a}} = \left( \lambda_3^{ab} + \lambda_4^{ab} + \kappa_A^2 \, \lambda_5^{ab} \right) \, \frac{v^2}{m_{A_{b}}^{2} - m_{A_{a}}^{2}} \, ,
\end{equation}
where we have introduced the $\kappa_A^2$ sign ($\kappa_R^2 = +1$ and $\kappa_I^2 = -1$).

\section{Neutrino masses}
\label{sec:numass}

\begin{figure}[h!]
\centering
\includegraphics[width=0.4\linewidth,keepaspectratio]{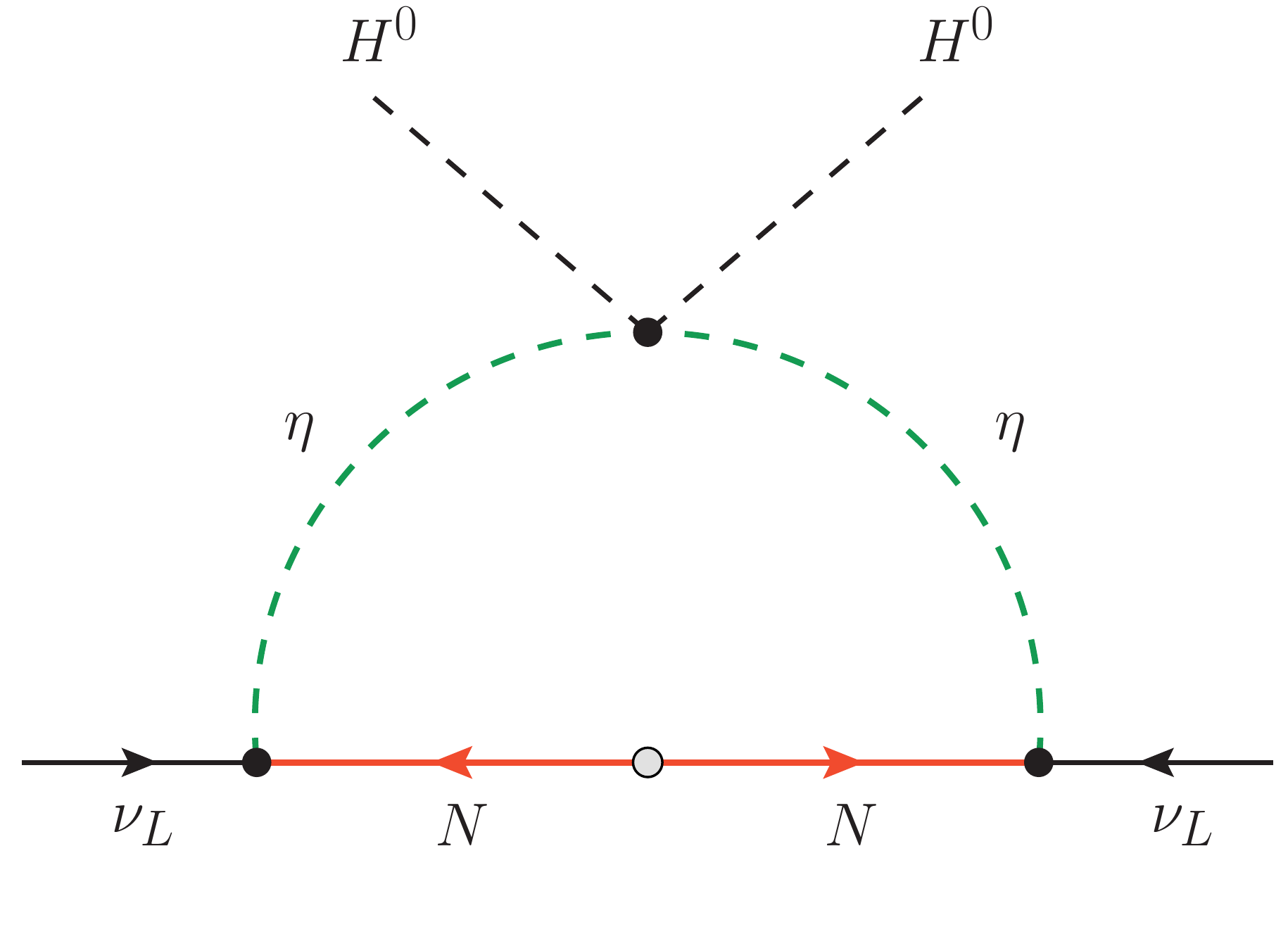}
\includegraphics[width=0.4\linewidth,keepaspectratio,trim= 0in -0.13in 0in 0in]{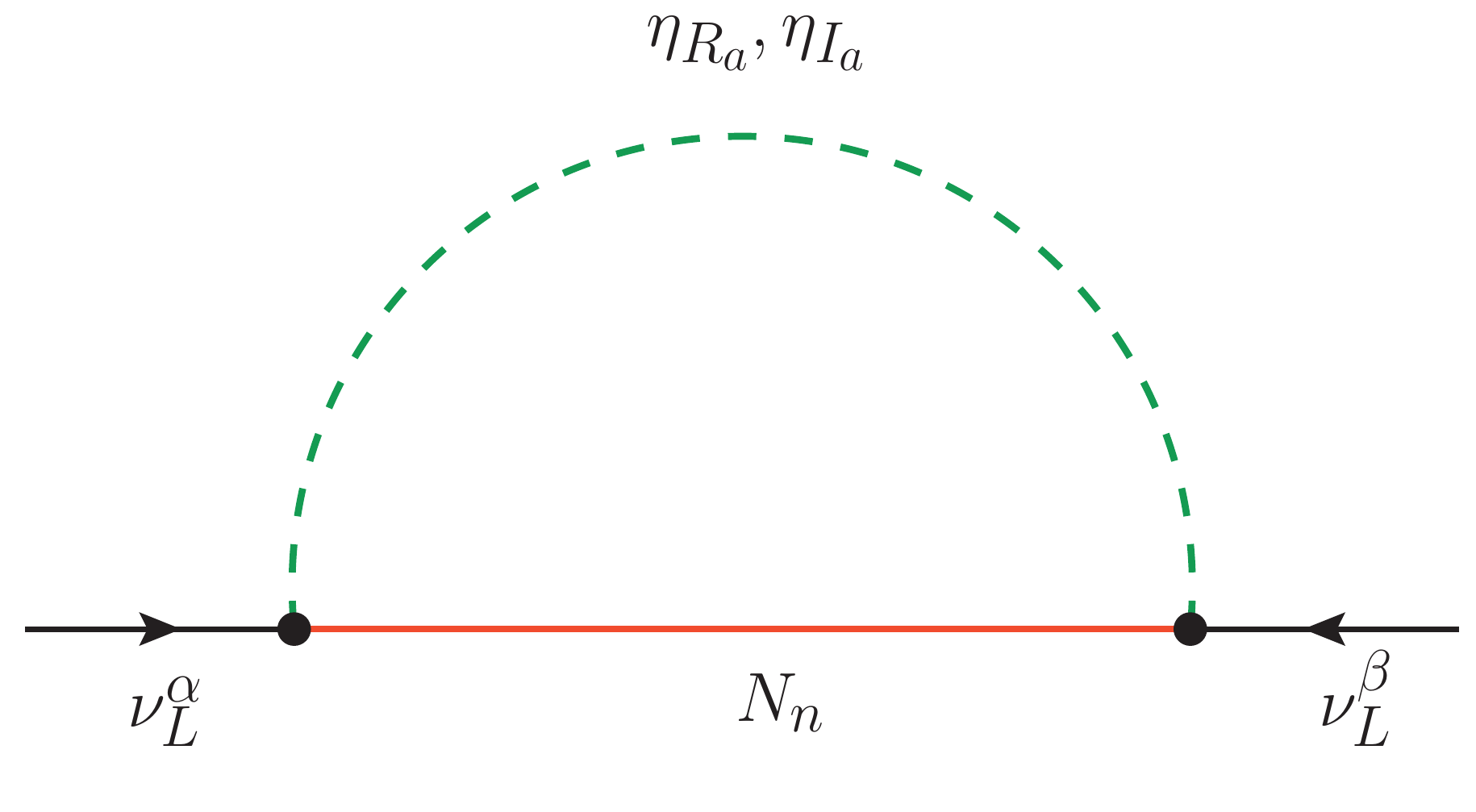}
\caption{ Feynman diagram for the neutrino mass generation with gauge eigenstates (left) and its equivalent with the physical mass eigenstates (right).
\label{fig:numass}}
\end{figure}


Given that it is assumed that $\eta_a$ do not get VEVs, neutrino masses are
forbidden at tree-level but induced at the 1-loop level, as shown in Fig.~\ref{fig:numass}, with several diagrams contributing to the neutrino mass matrix. Then, one can write $\left(m_\nu\right)_{\alpha \beta} = \sum_{A,a,n} \left(m_\nu^A\right)_{\alpha \beta}^{a n}$, where $\left(m_\nu^A\right)_{\alpha \beta}^{a n}$ is the contribution
to the neutrino mass matrix generated by the $N_n - \eta_{A_a}$ loop, and it is given by
\begin{equation} \label{eq:integral}
- i \left( m_\nu^A \right)_{\alpha \beta}^{a n} = C^A_{n a \alpha} \, \int \frac{\text{d}^D k}{\left( 2 \pi \right)^D} \frac{i}{k^2 - m^2_{A_a}} \frac{i \left( \slashed{k} + M_{N_n} \right)}{k^2 - M^2_{N_n}} \ C^A_{n a \beta} \, .
\end{equation}
Here $D = 4 - \varepsilon$, and is the number of space-time dimensions. $k$ is the momentum running
inside the loop and the external neutrinos are considered at rest. Notice that the term proportional to $\slashed{k}$ is not contributing because it is odd. Finally, $C^A_{n a \alpha}$ represents the $N_n - \eta_{A_a} - \nu_L^\alpha$ coupling and is given by
\begin{equation} \label{eq:Netanucoup}
C^A_{n a \alpha} = i \, \frac{\kappa_A}{\sqrt{2}} \, \sum_b \, \left( V_A \right)^*_{b a} \, y_{n b \alpha} \, ,
\end{equation}
with $\kappa_R = 1$ and $\kappa_I = i$. Note that we assume that the parameters in the scalar potential are real and the complex conjugation in $V_A$ can be dropped in the following. Now, replacing Eq.~\eqref{eq:Netanucoup} into Eq.~\eqref{eq:integral} and introducing the Passarino-Veltman loop function $B_0$~\cite{Passarino:1978jh},
\begin{equation}
B_0 \left( 0, m_{A_a}^2, M_{N_n}^2  \right) = \Delta_\varepsilon + 1 - \frac{m_{A_a}^2 \log m_{A_a}^2 - M_{N_n}^2 \log M_{N_n}^2}{m_{A_a}^2 - M_{N_n}^2},
\end{equation}
the neutrino mass matrix turns into
\begin{equation} \label{eq:mnu2}
\boxed{
\left( m_\nu \right)_{\alpha \beta} = - \frac{1}{32 \pi^2} \sum_{A, a, b, c, n} \, M_{N_n} \, \kappa_A^2 \left( V_A \right)_{b a} \, \left( V_A \right)_{c a} \, y_{n b \alpha} \, y_{n c \beta} \, B_0(0, m_{A_a}^2, M_{N_n}) \, , 
}
\end{equation}
where the divergent pieces, $\Delta_\varepsilon$, cancel exactly. Eq.~\eqref{eq:mnu2} constitutes the main result for the 1-loop neutrino mass matrix and it is a simple analytical expression for the neutrino mass matrix, but only in appearance. It involves a product of $V_A$ matrices and loop functions, both depending on the scalar potential parameters in a complicated way. Therefore, it is better to work under the assumptions in Eq.~\eqref{eq:assumptions} and derive an approximate expression for the neutrino mass matrix. It will be valid for small $\lambda_5^{a b}$ values and
small scalar mixing angles and the details of its derivation can be found in~\cite{Escribano:2020iqq}. Then, the approximate neutrino mass matrix can be expanded both in powers of $\lambda_5$ and also in powers of the small parameter
\begin{equation}
  s_{ab} = \frac{1}{2} \left(\lambda_3^{aa}+\lambda_4^{aa} \right) \frac{v^2}{2} \, ,
\end{equation}
which is defined for $a \neq b$ and it is nothing but $\sin \theta_A^{ab}$ at 0th order in $\lambda_5$. This way, the approximate matrix can be written as
\begin{equation} \label{eq:mnu3}
\boxed{
  \left( m_\nu \right)_{\alpha \beta} = \frac{v^2}{32 \pi^2} \sum_{n, a, b} \frac{y_{n a \alpha} \, y_{n b \beta}}{M_{N_n}} \, \Gamma_{abn} + \mathcal{O} \left( \lambda_5^2 \right) + \mathcal{O} \left( \lambda_5 \, s^2 \right) \, ,
}
\end{equation}
where, for simplicity, we have defined the dimensionless tensor
\begin{equation} \label{eq:Gamma}
\Gamma_{abn} = \delta_{ab} \, \lambda_5^{aa} \, f_{an} - (1 - \delta_{ab}) \left[ \left( \lambda_5^{aa} \, f_{an} - \lambda_5^{bb} \, f_{bn} \right) \, s_{ab} - \frac{M^2_{N_n}}{m_b^2 - m_a^2} \, \lambda_5^{ab} \ g_{abn} \right]
\end{equation}
and the loop functions
\begin{equation}
  f_{an} = \frac{M_{N_n}^{2}}{m_a^2 - M_{N_n}^{2}} + \frac{M_{N_n}^{4}}{\left(m_a^2 - M_{N_n}^{2}\right)^{2}} L_{an} \quad , \quad g_{abn} = \frac{m_a^2}{m_a^2 - M_{N_n}^2} L_{an} - \frac{m_b^2}{m_b^2 - M_{N_n}^2} L_{bn} \, ,
\end{equation}
%
%
where $L_{an} = \log \frac{M_{N_n}^{2}}{m_a^2}$. Also, $m_a$ are the scalar masses at 0th order in
$\lambda_5^{aa}$, that is,
\begin{equation}
  m_{a}^{2} = \left(m_{\eta}^{2}\right)_{a a} + \left( \lambda_{3}^{a
    a} + \lambda_{4}^{a a} \right) \, \frac{v^2}{2} \, .
\end{equation}
Finally, eq.~\eqref{eq:mnu3} reproduces the
neutrino mass matrix in very good approximation and it is valid for any
$n_N$ and $n_\eta$ values. 

\section{High-energy behavior}
\label{sec:running}

A crucial point for the consistency of the Scotogenic setup is the conservation of the $\mathbb{Z}_2$ parity. If this symmetry is absent at some energy scale,
neutrinos would acquire masses at tree-level and we will lose the DM candidate
stability. The study of the conservation of the $\mathbb{Z}_2$ symmetry at high energies was initiated in~\cite{Merle:2015gea}, where it was pointed out that the RGE flow might alter the shape of the scalar potential at high energies, eventually breaking the symmetry. Actually, later
works~\cite{Merle:2015ica,Lindner:2016kqk} showed that this problem is present in large portions of the parameter space. 

We can understand some features of the behavior of the model at high energies by examining the 1-loop $\beta$ function for the $m_\eta^2$ parameter,
\vspace{-0.1cm}
\begin{equation}
  \begin{split}
    & \left(\beta_{m^2_\eta}\right)_{ab} = - \frac{9}{10} \, g_1^2 \, \left(m^2_\eta\right)_{a b} - \frac{9}{2} \, g_2^2 \, \left(m^2_\eta\right)_{a b} + \sum_{c,d=1}^{\neta} \left[ 4 \, \lambda_2^{abcd} \left(m^2_\eta\right)_{dc} + 2 \, \lambda_2^{acdb} \left(m^2_\eta\right)_{cd} \right] \\
    & + \left[  4 \, \lambda_3^{ab} + 2 \, \lambda_4^{ab}  \right] \, m^2_H + \left(m^2_\eta\right)_{ab} \sum_{n = 1}^{\nN} \sum_{\alpha = 1}^3 \left( \left| y _{n a \alpha} \right|^2 +  \left| y _{n b \alpha} \right|^2  \right) - 4 \, \Tr \left[ y_a^\dagger M_N^\ast M_N y_b \right] \, . 
  \end{split}
  \label{eq:RGEmeta2}
\end{equation}
It turns out that the last term, with its negative contribution, plays a crucial role in the breaking of the symmetry and we will refer to it as \textit{the trace term}. For the
standard Scotogenic model, it was first noted in~\cite{Merle:2015gea} that in case of large Yukawas, which is equivalent to $\lambda_5 \ll 1$, and $M_N^2 \gtrsim m_\eta^2$, the running is dominated by the trace term, driving it to negative values and eventually breaking the
symmetry. This happens once $m_\eta^2 < 0$, which induces
a minimum of the scalar potential for which $\langle \eta \rangle \ne
0$. We expect the same behavior in the general Scotogenic model as well. 

In order to explore the scalar potential of the model at high energies, we solve the full set of RGEs numerically for one specific variant of the general Scotogenic model: the {\bf $\boldsymbol{(1,3)}$ model}, with one singlet fermion and three inert doublets. What we do is to set all the parameters of the model at the electroweak scale, but always checking that the scalar potential is bounded from below at this scale. A detailed discussion on how we
check the boundedness from below is shown in~\cite{Escribano:2020iqq}. Finally, the
neutrino squared mass differences and the mixing angles that have been
measured in neutrino oscillation experiments must be accommodated by appropriately fixing the
Yukawa couplings. This can be done, as explained
in~\cite{Toma:2013zsa,Vicente:2014wga,Cordero-Carrion:2018xre,Cordero-Carrion:2019qtu}, by means of an adapted Casas-Ibarra parametrization~\cite{Casas:2001sr}.

\begin{figure}[tb!]
\centering
\includegraphics[width=0.44\linewidth,keepaspectratio]{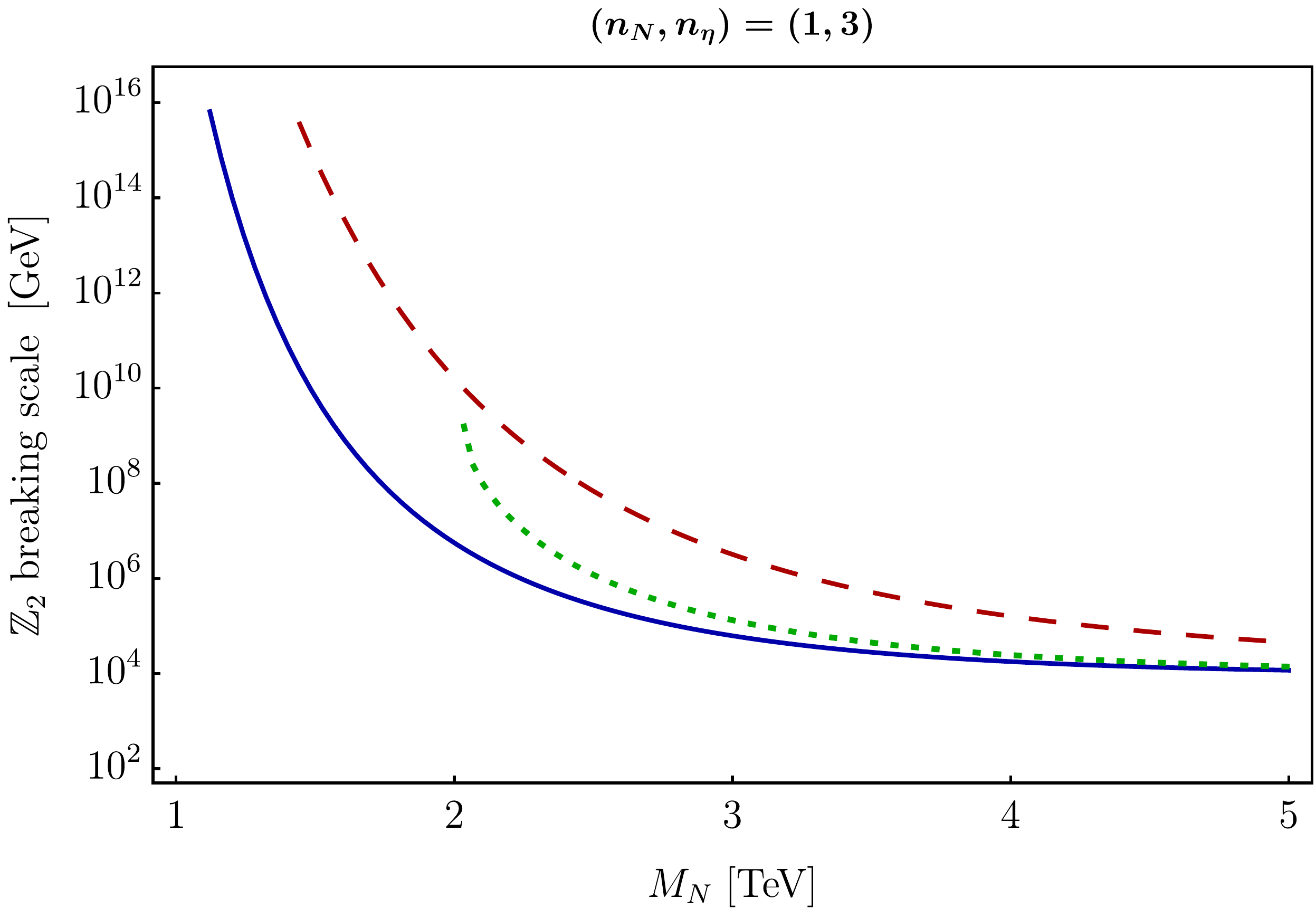}
\includegraphics[width=0.44\linewidth,keepaspectratio]{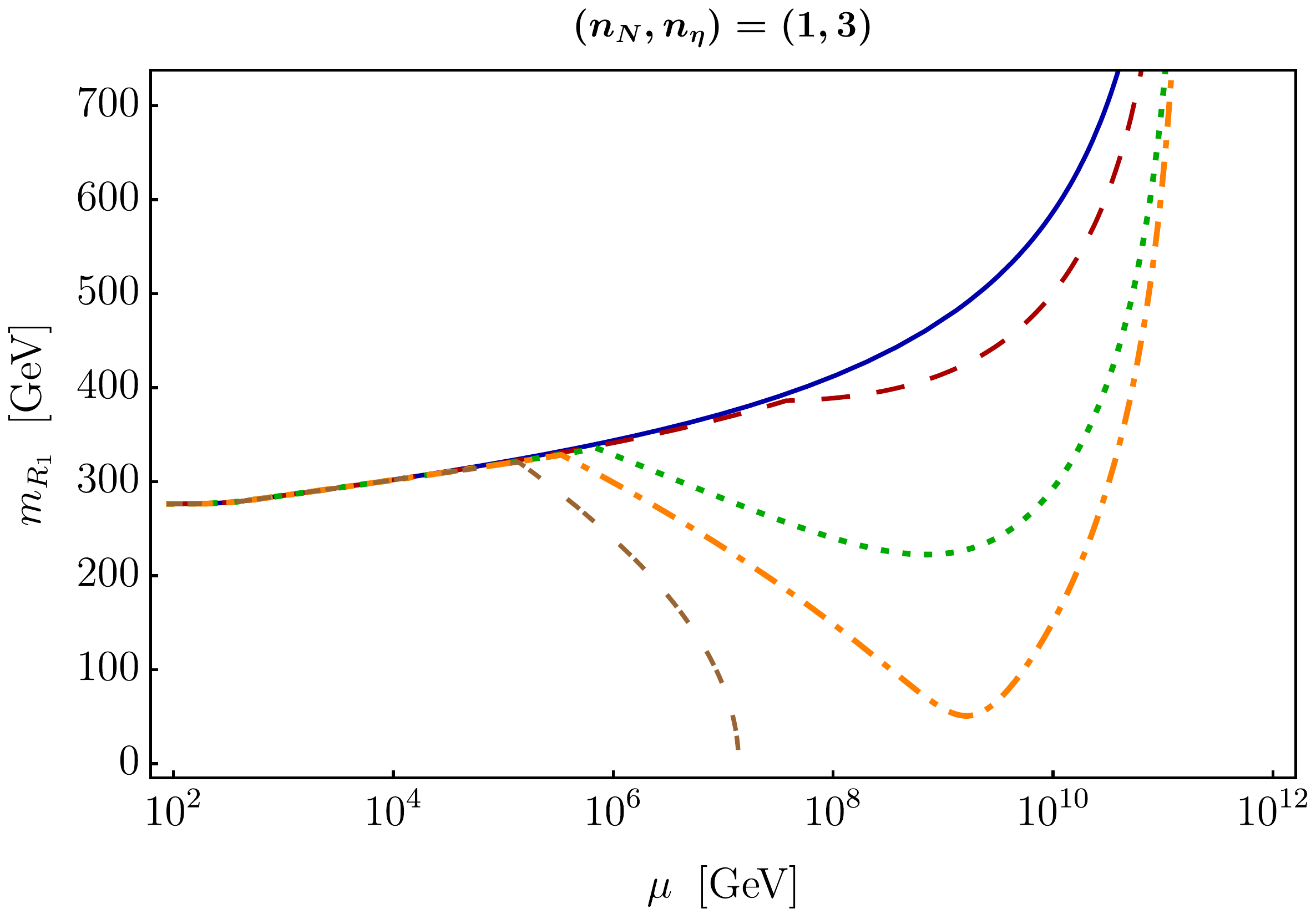}
\caption{On the left-hand side the $\mathbb{Z}_2$ breaking scale is shown as a function of the singlet
  fermion mass $M_N$ in the $(1,3)$ Scotogenic model for three
  different scenarios: $\lambda_2^{aaaa} = \lambda_3^{aa} =
  \lambda_4^{aa} = 0.1$ and $m_\eta^2 = \left( 200^2, 300^2, 400^2
  \right)$~GeV$^2$ (blue), $\lambda_2^{aaaa} =
  \lambda_2^{aabb} = \lambda_3^{aa} = \lambda_4^{aa} = 0.3$ and
  $m_\eta^2 = \left( 200^2, 300^2, 400^2 \right)$~GeV$^2$ (green,
  dotted), and $\lambda_2^{aaaa} = \lambda_3^{aa} =
  \lambda_4^{aa} = 0.1$ and $m_\eta^2 = \left( 200^2, 600^2, 800^2
  \right)$~GeV$^2$ (red, dashed). The remaining scalar parameters are set to zero and $\lambda_5^{aa} = 10^{-9}$ in all the cases. On the right-hand side the evolution of the lightest scalar mass $m_{R_1}$ as a function
  of the energy scale $\mu$ in the $(1,3)$ Scotogenic model is shown. Here the values for the previous green dotted curve are used for the scalar parameters, and
  $M_N$ takes the values $1$~TeV (blue), $1.5$~TeV (red, dashed),
  $1.9$~TeV (green, dotted), $2.025$~TeV (orange, dash-dotted) and
  $2.2$~TeV (brown, double dashed).
  \label{fig:Z2MN}}
\end{figure}

Now we are in position to explore the behavior of the model at high energies. First, as already seen in~\eqref{eq:RGEmeta2}, the singlet fermion mass $M_N$ is responsible of the breaking of the symmetry by driving the scalar masses towards negative values through the trace term. On the left-hand side of Fig.~\ref{fig:Z2MN}, we show the scale at which the symmetry gets broken as a function of the singlet mass for several
parameter sets. For the blue and red lines we have used moderate
values for the quartic couplings, $\lambda_2^{aaaa} = \lambda_3^{aa} =
\lambda_4^{aa} = 0.1$, while for the green line increased and
additional quartics have been used, $\lambda_2^{aaaa} = \lambda_2^{aabb} =
\lambda_3^{aa} = \lambda_4^{aa} = 0.3$. Regarding the $\lambda_5$ matrix, it is
taken to be diagonal, with $\lambda_5^{aa} = 10^{-9}$. 
As we could expect, the effect of the trace term becomes stronger for larger singlet masses, therefore, the $\mathbb{Z}_2$ breaking scale decreases and this generic behavior can be found in large regions of the parameter space. Notice that for $M_N \lesssim 2$ TeV the $\mathbb{Z}_2$ symmetry is not getting broken for the green curve, as will be understand in a moment. 

The right-hand side of Fig.~\ref{fig:Z2MN} shows how the lightest scalar mass
$m_{R_1}$ evolves with the energy scale for the parameter set
corresponding to the green curve of the previous plot. The different curves
have been obtained for several values of $M_N$. For $M_N = 2.2$ TeV one can observe that the scalar mass reaches zero, breaking the symmetry at $\mu \simeq 10^7$ GeV,
in agreement with the plot on the left. Then, the lower the singlet mass is, the faster $m_{R_1}$ increases to the infinity, even if it is initially decreased due to the effect of the trace term. The reason for this is that a Landau pole appears in the $\lambda_2$ quartic couplings, making them grow with the energy until they completely dominate the
$\beta$ function with a positive contribution. This is clearly seen in Eq.~\eqref{eq:RGEmeta2}. Finally, we note that this Landau pole is present, for many choices of the parameters, at energies well above the $\mathbb{Z}_2$ breaking scale.

\section{Summary and discussion}
\label{sec:conclusions}

In this work we have presented a generalization of the Scotogenic model to arbitrary numbers of the generations of the singlet fermions and the inert
doublets. We have computed the 1-loop neutrino mass matrix in the
general version of the model, both exactly and approximately, and we have studied the high-energy
behavior of a specific variant with $(n_N,n_\eta) = (1,3)$. 
We conclude that the features of the original Scotogenic model are kept in the multi-scalar version as well. Furthermore, novel phenomenological signatures might also exist. The couplings of the $\eta$ scalars to the gauge bosons allow them to be produced at the LHC. Therefore, exotic signatures might be possible in versions with many $\eta$ generations. For instance, the production of the heaviest $\eta$ doublets would initiate cascade decays, leading to striking
multilepton signatures with missing energy due to the production
of the lightest $\mathbb{Z}_2$-odd state. Finally, the presence of additional scalars might affect the dark matter production rates in the early Universe. 

\section*{Acknowledgements}

This article is based on the talk given at TAUP2021. The original work \cite{Escribano:2020iqq} was done in collaboration with Mario Reig and Avelino Vicente. Work supported by the Spanish grants
PID2020-113775GB-I00 (AEI / 10.13039/501100011033), PROMETEO/2018/165 (Generalitat Valenciana) and by the FPI grant PRE2018-084599.

\section*{References}
\bibliographystyle{iopart-num}
\bibliography{refs}

\end{document}